\documentclass[letterpaper, 10 pt, conference]{ieeeconf}  % Comment this line out if you need a4paper

\IEEEoverridecommandlockouts                              % This command is only needed if 
\usepackage[scheme=plain]{ctex}

\usepackage[breaklinks,colorlinks,linkcolor=black,citecolor=black,urlcolor=black]{hyperref}
\usepackage{amsmath, amssymb, amsthm}

\usepackage{tabularx}
\usepackage{algorithm}
\usepackage{algorithmicx}
\usepackage{algpseudocode}
\usepackage{booktabs}
\usepackage{mathrsfs}
\usepackage{tikz}
\usepackage{tikz-cd}
\usetikzlibrary{shapes,arrows,positioning,fit, shapes.geometric, arrows.meta, positioning, calc, decorations.pathreplacing,patterns}

\usepackage{pgfplots}
\usepackage{pgfkeys}
\pgfplotsset{compat=1.18}

\hypersetup{
	colorlinks=true,
	linkcolor=blue,
	citecolor=green,
	urlcolor=red,
}

\newtheorem{definition}{Definition}[section]
\newtheorem{theorem}{Theorem}[section]
\newtheorem{lemma}[theorem]{Lemma}

\begin{document}
	
	\title{Dual-Tape Perspective and Generator Independence: The Algebraic Foundation of Real Boolean Turing Machines}

	\author{Jingwen Zheng$^{1}$ and Bojin Zheng$^{2}$ and  Weiwu Wang$^{3}$%
		%	\thanks{*This work was not supported by any organization}% <-this % stops a space
		\thanks{$^{1}$ Jingwen Zheng,  College of Cyber Science and Engineering, Wuhan University, Wuhan, 430072 China
			{\tt\small zhengjw@whu.edu.cn}
		}
		\thanks{$^{2}$ Bojin Zheng, College of Computer Science,  South-Central Minzu University, Wuhan, 430074 China
	       {\tt\small zhengbj@mail.scuec.edu.cn}
        }
		\thanks{$^{3}$ Weiwu Wang, Information Technology Department, China Development Bank, Beijing, 100033 China
			{\tt\small hiwangww@163.com}
		}
}

	\maketitle
	
	\begin{abstract}
		The Complex Boolean Turing Machine (CBTM) characterizes non-deterministic computation using the abstract generator $\alpha$, but the abstractness of $\alpha$ makes it difficult to understand intuitively. In this paper, by concretizing $\alpha$ as the algebraic number $\sqrt{2}$, we introduce the \textbf{Real Boolean Turing Machine (RBTM)} and propose the \textbf{dual-tape perspective}—decomposing each tape into a real tape (storing rational coefficients $a$) and an imaginary tape (storing irrational coefficients $b$). The ``1''s on the imaginary tape intuitively mark the locations of ``new dimensions,'' laying a physical foundation for subsequent dynamic dimension tracking. More importantly, we prove the \textbf{Generator Independence Theorem}: computational power is independent of the specific choice of generator—whether using $\sqrt{2}$, $\sqrt{3}$, or the imaginary unit $i$, the corresponding automata are isomorphic. This reveals that the essence of non-determinism lies in the fact of ``introducing a new element incommensurable with the base field,'' rather than the algebraic identity of the generator. Furthermore, we introduce the \textbf{generator extraction operator} and analyze its limitations within a static framework, highlighting the necessity of introducing a dynamic IVM. The RBTM serves both as a visualized instance of the CBTM and as a bridge to the subsequent dynamic dimension tracking of the Imaginary Part Verifier (IVM).
	\end{abstract}
	
	\section{Introduction}
	\label{sec:intro}
	
	\subsection{Review: CBTM and the Abstract $\alpha$}
	In the second paper of this series~\cite{cbtm}, we proposed the Complex Boolean Turing Machine (CBTM), elevating computational symbols to algebraic elements in $\operatorname{GF}(4)$, giving each operation explicit mathematical semantics. The core insight of the CBTM is that \textbf{non-deterministic computation corresponds to algebraic field extension}—when reading a symbol $\alpha$ representing a ``new dimension,'' the computation must branch into two paths. However, as an abstract generator, $\alpha$ lacks intuitive geometric or physical meaning. This raises two natural questions:
	\begin{itemize}
		\item Can we concretize $\alpha$ as a familiar algebraic number (such as $\sqrt{2}$), making the ``new dimension'' tangible?
		\item If we change the generator (for example, replacing $\sqrt{2}$ with $\sqrt{3}$ or $i$), does the computational power change?
	\end{itemize}
	
	\subsection{Question I: How to Visualize ``Dimension''?}
	To answer the first question, we choose the most classic quadratic irrational $\sqrt{2}$ as the generator, representing each symbol as $a+b\sqrt{2}$ ($a,b\in\{0,1\}$). In this way, the algebraic semantics of symbols become concrete: the real part $a$ represents the traditional Boolean value, and the imaginary coefficient $b$ indicates whether the new dimension $\sqrt{2}$ is involved. More importantly, we can decompose abstract algebraic symbols into two physically independent Boolean tapes—the \textbf{dual-tape perspective}. This decomposition will make the concept of ``dimension'' intuitively visible for the first time: every ``1'' on the imaginary tape marks the presence of a new dimension.
	
	\subsection{Question II: Does the Choice of Generator Affect Computational Power?}
	The second question is more profound. If we replace $\sqrt{2}$ with $\sqrt{3}$ or $i$, does the computational power change? Intuitively, as long as the new number is ``incommensurable'' with the old numbers (i.e., not in the same field), they should play the same role. But we need a rigorous mathematical proof. If the proof holds, it indicates that the CBTM framework captures the essence of non-determinism independent of the specific algebraic implementation; conversely, if computational power changes with the generator, the foundation of the framework would be shaken.
	
	\subsection{Contributions of This Paper}
	This paper provides a complete answer to the above questions by introducing the \textbf{Real Boolean Turing Machine (RBTM)}:
	\begin{enumerate}
		\item Concretize $\alpha$ as $\sqrt{2}$ and introduce the \textbf{dual-tape perspective}, making the ``new dimension'' visualizable (Section \ref{sec:dual-tape}).
		\item Prove the computational equivalence between RBTM and CBTM, ensuring that concretization does not alter computational power (Section \ref{sec:rbtm}).
		\item Prove the \textbf{Generator Independence Theorem}: For any two non-rational algebraic numbers $\gamma$ and $\delta$, the RBTM based on $\gamma$ and the RBTM based on $\delta$ are automaton-isomorphic, and thus have identical computational power (Section \ref{sec:independence}).
		\item Introduce the \textbf{generator extraction operator} and analyze its static limitations, revealing the necessity of transitioning from RBTM to a dynamic IVM (Section \ref{sec:generator-operator}).
	\end{enumerate}
	The RBTM plays a connecting role in the series: it is both a visualized instance of the CBTM and provides the physical foundation for the dynamic dimension tracking of the subsequent Imaginary Part Verifier (IVM)~\cite{ivm}.
	
	\section{Dual-Tape Perspective — Visualizing Dimensions}
	\label{sec:dual-tape}
	
	\subsection{From $\alpha$ to $\sqrt{2}$}
	The generator $\alpha$ in CBTM satisfies $\alpha^2=\alpha+1$, an abstract symbol. To obtain intuitive geometric meaning, we replace it with the concrete algebraic number $\sqrt{2}$. $\sqrt{2}$ is incommensurable with the rational number field $\mathbb{Q}$, which is precisely the property required for a ``new dimension.'' Thus, each symbol can be represented as
	\[
	\sigma = a + b\sqrt{2},\quad a,b\in\{0,1\}.
	\]
	The real part $a$ encodes the traditional Boolean value, and the imaginary coefficient $b$ indicates whether $\sqrt{2}$ is involved.
	
	\subsection{Dual-Tape Decomposition}
	Physically decompose a single tape into two parallel Boolean tapes:
	\begin{itemize}
		\item \textbf{Real Tape}: stores the real coefficient $a$ at each position;
		\item \textbf{Imaginary Tape}: stores the imaginary coefficient $b$ at each position.
	\end{itemize}
	This decomposition establishes a natural isomorphism $\mathcal{CB}\cong\{0,1\}\times\{0,1\}$ and is structure-preserving.
	
	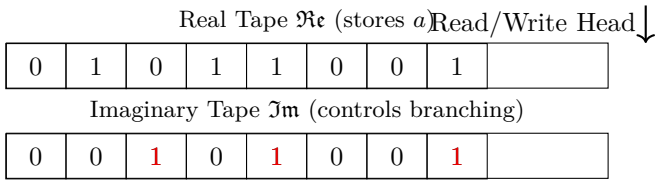
\begin{figure}[htbp]
		\centering
		\begin{tikzpicture}[
			tape/.style={draw, minimum width=8cm, minimum height=0.6cm, inner sep=0pt},
			cell/.style={draw, minimum width=0.8cm, minimum height=0.6cm, outer sep=0pt},
			label/.style={font=\small}
			]
			% Real tape
			\node[tape] (real-tape) at (0,0) {};
			\foreach \x in {0,1,...,7} {
				\node[cell] at ($(real-tape.south west) + (\x*0.8+0.4,0.3)$) {};
			}
			\node[label, above] at (real-tape.north) {Real Tape $\mathfrak{Re}$ (stores $a$)};
			\foreach \x/\val in {0/0,1/1,2/0,3/1,4/1,5/0,6/0,7/1} {
				\node at ($(real-tape.south west)+(0.4+\x*0.8,0.3)$) {\val};
			}
			% Imaginary tape
			\node[tape] (imag-tape) at (0,-1.2) {};
			\foreach \x in {0,1,...,7} {
				\node[cell] at ($(imag-tape.south west) + (\x*0.8+0.4,0.3)$) {};
			}
			\node[label, above] at (imag-tape.north) {Imaginary Tape $\mathfrak{Im}$ (controls branching)};
			\foreach \x/\val in {0/0,1/0,2/1,3/0,4/1,5/0,6/0,7/1} {
				\node at ($(imag-tape.south west)+(0.4+\x*0.8,0.3)$) {\val};
			}
			% Highlight positions where imaginary part is 1 in red
			\node[red, thick] at ($(imag-tape.south west)+(2.0,0.3)$) {1};
			\node[red, thick] at ($(imag-tape.south west)+(3.6,0.3)$) {1};
			\node[red, thick] at ($(imag-tape.south west)+(6.0,0.3)$) {1};
			% Read/write head
			\draw[->, thick] (4.5,0.8) -- (4.5,0.3) node[midway, left] {Read/Write Head};
		\end{tikzpicture}
		\caption{Dual-tape perspective: the real tape stores traditional data; the red ``1''s on the imaginary tape mark positions of new dimensions, triggering non-deterministic branching.}
		\label{fig:dual-tape}
	\end{figure}
	
	The dual-tape perspective provides powerful intuition:
	\begin{itemize}
		\item The real tape performs traditional, deterministic Boolean computation;
		\item The imaginary tape controls non-determinism: when a position on the imaginary tape is 1, it means ``a new dimension is embedded here,'' and the machine must branch, simultaneously exploring both possibilities of including and excluding that dimension.
	\end{itemize}
	
	\subsection{Resolution of the Field Non-Isomorphism Problem via the Dual-Tape Perspective}
	The algebraic foundation of the CBTM rests on the finite field $\operatorname{GF}(4)$ (characteristic 2), while the RBTM is based on the infinite field $\mathbb{Q}(\sqrt{2})$ (characteristic 0). As algebraic structures, these two fields are non-isomorphic, one finite and one infinite. However, under the dual-tape perspective, each symbol is decomposed into two Boolean components (real and imaginary), and computational behavior is entirely determined by the evolution of these two Boolean components, independent of the specific algebraic properties of the generator. Therefore, despite the structural differences between $\operatorname{GF}(4)$ and $\mathbb{Q}(\sqrt{2})$, the automata they induce at the Boolean projection level are \textbf{automaton-isomorphic}. This is a corollary of the Generator Independence Theorem: different algebraic representations can yield exactly the same computational behavior. The dual-tape perspective eliminates differences in the underlying field structure by ``projecting'' algebraic symbols onto the Boolean level, allowing us to focus on the essence of computation—the presence or absence of new dimensions, rather than the algebraic identity of those dimensions.
	
	\section{Real Boolean Turing Machine (RBTM) — A Concrete Instance of CBTM}
	\label{sec:rbtm}
	
	\subsection{Formal Definition}
	\begin{definition}[RBTM]
		A Real Boolean Turing Machine is defined as a seven-tuple
		\[
		M_{\mathrm{RBTM}} = (Q,\Sigma,\Gamma,\delta,q_0,F,\epsilon),
		\]
		where:
		\begin{itemize}
			\item $Q$: a finite set of states;
			\item $\Sigma \subseteq \mathbb{Q}(\sqrt{2})$: the input alphabet, each symbol of the form $a+b\sqrt{2}$, $a,b\in\{0,1\}$;
			\item $\Gamma = \Sigma \cup \{\#\}$, where $\#$ is the blank symbol, with projections defined as $\mathfrak{Re}(\#)=\mathfrak{Im}(\#)=0$;
			\item $\delta: Q\times\Gamma \to \mathcal{P}(Q\times\Gamma\times\{L,R\})$: the transition function, satisfying the Branch Triggering Axiom and the Projection Constraint Axiom;
			\item $q_0\in Q$: the initial state;
			\item $F\subseteq Q$: the set of accepting states;
			\item $\epsilon\in\mathbb{Q}^+$: the branching threshold (typically $\epsilon=0.5$).
		\end{itemize}
		The Branch Triggering Axiom and the Projection Constraint Axiom are exactly the same as in the CBTM, with $\alpha$ in CBTM replaced by $\sqrt{2}$.
	\end{definition}
	
	\subsection{Equivalence with CBTM}
	\begin{theorem}[Equivalence of RBTM and CBTM]
		There exist polynomial-time mutual simulations such that $\mathcal{L}(\mathrm{RBTM})=\mathcal{L}(\mathrm{CBTM})$, and the complexity classes coincide: $\mathbf{P}_{rb}=\mathbf{P}_{cb}=\mathbf{P}$, $\mathbf{NP}_{rb}=\mathbf{NP}_{cb}=\mathbf{NP}$.
	\end{theorem}
	\begin{proof}
		Define the symbol mapping $\psi:\mathcal{CB}\to\mathbb{Q}(\sqrt{2})$ as:
		\[
		\psi(0)=0,\quad \psi(1)=1,\quad \psi(\alpha)=\sqrt{2},\quad \psi(\beta)=1+\sqrt{2}.
		\]
		It is easy to verify that $\psi$ preserves projection values: $\mathfrak{Re}_{\mathrm{CBTM}}(\sigma)=\mathfrak{Re}_{\mathrm{RBTM}}(\psi(\sigma))$, $\mathfrak{Im}_{\mathrm{CBTM}}(\sigma)=\mathfrak{Im}_{\mathrm{RBTM}}(\psi(\sigma))$. Map each configuration of the CBTM symbolwise via $\psi$ to obtain a configuration of the RBTM; conversely, use the inverse mapping to map RBTM configurations back to CBTM. Since the transition function is determined by projection values, this mapping preserves the transition relation, with $O(1)$ overhead per step. Thus the two models are polynomial-time equivalent.
	\end{proof}
	
	\subsection{Computation Example}
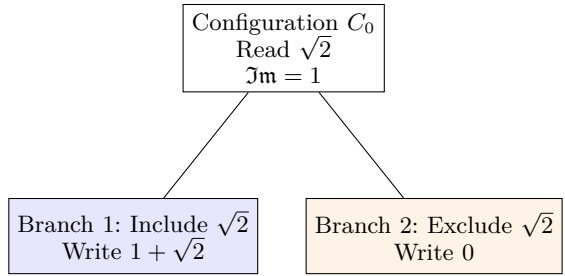
\begin{figure}[htbp]
		\centering
		\begin{tikzpicture}[
			level distance=2.5cm,
			level 1/.style={sibling distance=4cm},
			node/.style={draw, rectangle, minimum width=2cm, minimum height=1cm, align=center, font=\small}
			]
			\node [node] {Configuration $C_0$\\Read $\sqrt{2}$\\$\mathfrak{Im}=1$}
			child { 
				node [node, fill=blue!10] {Branch 1: Include $\sqrt{2}$\\Write $1+\sqrt{2}$}
			}
			child {
				node [node, fill=orange!10] {Branch 2: Exclude $\sqrt{2}$\\Write $0$}
			};
		\end{tikzpicture}
		\caption{RBTM single-step branching example: imaginary part 1 triggers a binary branch, corresponding to inclusion or exclusion of the new dimension.}
		\label{fig:single-branch}
	\end{figure}

	Consider an RBTM processing a single symbol $\sqrt{2}$ (i.e., $\sigma=0+1\cdot\sqrt{2}$). Under the dual-tape perspective, the real tape has 0 at that position and the imaginary tape has 1. The imaginary part being 1 triggers branching: the left branch corresponds to ``include $\sqrt{2}$,'' writing $1+\sqrt{2}$ (real part 1, imaginary part 1); the right branch corresponds to ``exclude $\sqrt{2}$,'' writing $0$ (real part 0, imaginary part 0). The computation tree is shown in Figure \ref{fig:single-branch}.
	
	\section{Generator Independence — The Essence of Non-Determinism}
	\label{sec:independence}
	
	\subsection{Formalizing the Problem}
	Let $\gamma$ and $\delta$ be any two non-rational algebraic numbers (e.g., $\sqrt{2}$ and $\sqrt{3}$, or $\sqrt{2}$ and $i$). Construct RBTMs $M_\gamma$ and $M_\delta$ using $\gamma$ and $\delta$ as generators, respectively. Question: Are $M_\gamma$ and $M_\delta$ computationally equivalent?
	
	\subsection{Key Observation}
	Branch triggering depends only on the Boolean value of the imaginary coefficient ($\mathfrak{Im}(\sigma)=b$), independent of the specific numeric value of the generator. Regardless of whether the generator is $\sqrt{2}$ or $i$, as long as the imaginary coefficient $b=1$, the machine must branch. Therefore, any difference in behavior between the two machines can only arise from algebraic operations on the symbols, but the projection operators extract only the rational coefficients $a,b$, which remain unchanged under generator replacement.
	
	\subsection{Consistency of Projection Values}
	\begin{lemma}[Consistency of Projection Values]
		For any symbol $\sigma=a+b\gamma$ ($a,b\in\mathbb{Q}$), its projections under generator $\gamma$ are $\mathfrak{Re}_\gamma(\sigma)=a$, $\mathfrak{Im}_\gamma(\sigma)=b$. If $\gamma$ is replaced by another generator $\delta$, yielding $\sigma'=a+b\delta$, then $\mathfrak{Re}_\delta(\sigma')=a$, $\mathfrak{Im}_\delta(\sigma')=b$. That is, projection values depend only on the rational coefficients $a,b$, not on the generator.
	\end{lemma}
	\begin{proof}
		Directly follows from the definition of the projection operators.
	\end{proof}
	
	\subsection{Constructing Automaton Isomorphism}
	\begin{definition}[Automaton Isomorphism]
		Two automata $M_1=(Q_1,\Sigma_1,\delta_1,q_{01},F_1)$ and $M_2=(Q_2,\Sigma_2,\delta_2,q_{02},F_2)$ are said to be isomorphic (denoted $M_1\cong M_2$) if there exists a bijection $\phi:Q_1\to Q_2$ satisfying:
		\begin{enumerate}
			\item $\phi(q_{01})=q_{02}$;
			\item $\phi(F_1)=F_2$;
			\item For every $q\in Q_1$ and $\sigma\in\Sigma_1$, $\phi(\delta_1(q,\sigma))=\delta_2(\phi(q),\sigma)$ (here we interpret the range of $\delta$ as the state set, ignoring written symbols and head movements, as these can be encoded in the state; for a rigorous treatment, a configuration isomorphism is required, but for simplicity we focus on the correspondence of state transitions; the isomorphism of tape symbols is guaranteed by the consistency of projection values).
		\end{enumerate}
	\end{definition}
	
	\begin{theorem}[Generator Independence]
		Let $\gamma$ and $\delta$ be any two non-rational algebraic numbers. Then the RBTM $M_\gamma$ based on $\gamma$ and the RBTM $M_\delta$ based on $\delta$ are automaton-isomorphic. In particular, they recognize the same language class and have the same complexity class.
	\end{theorem}
	\begin{proof}
		First, construct a bijection between the alphabets. Denote $\Gamma_\gamma = \{0,1,\gamma,1+\gamma\}$, $\Gamma_\delta = \{0,1,\delta,1+\delta\}$. Define the mapping $\psi:\Gamma_\gamma\to\Gamma_\delta$ as
		\[
		\psi(a+b\gamma)=a+b\delta,\qquad a,b\in\{0,1\}.
		\]
		That is, $\psi(0)=0,\;\psi(1)=1,\;\psi(\gamma)=\delta,\;\psi(1+\gamma)=1+\delta$. Clearly $\psi$ is bijective and preserves projection values:
		\[
		\mathfrak{Re}_\delta(\psi(\sigma)) = a = \mathfrak{Re}_\gamma(\sigma)
		\],
		\[ 
		\mathfrak{Im}_\delta(\psi(\sigma)) = b = \mathfrak{Im}_\gamma(\sigma).
		\]
		
		Take the identity map $\iota:Q_\gamma\to Q_\delta$ for the states (the state sets of the two machines can be considered identical, as they depend only on the control logic, not on the specific generator value). We now verify that the transition relation is preserved under $(\iota,\psi)$.
		
		Consider an arbitrary transition $(q',\gamma_w,d)\in\delta_\gamma(q,\sigma)$ in $M_\gamma$, where $\sigma=a+b\gamma$. According to the Projection Constraint Axiom, there exist Boolean functions $f_{\mathfrak{Re}},f_{\mathfrak{Im}}$ such that the written symbol $\gamma_w$ satisfies:
		\[
		\gamma_w = f_{\mathfrak{Re}}(q,\mathfrak{Re}_\gamma(\sigma)) + \gamma\cdot f_{\mathfrak{Im}}(q,\mathfrak{Im}_\gamma(\sigma)).
		\]
		In $M_\delta$, the read symbol is $\psi(\sigma)=a+b\delta$, whose projection values are the same as those of $\sigma$. By the same Boolean functions $f_{\mathfrak{Re}},f_{\mathfrak{Im}}$, according to the Projection Constraint Axiom, $M_\delta$ in state $\iota(q)=q$ must have a transition $(q',\delta_w,d)$ such that
		\[
		\delta_w = f_{\mathfrak{Re}}(q,\mathfrak{Re}_\delta(\psi(\sigma))) + \delta\cdot f_{\mathfrak{Im}}(q,\mathfrak{Im}_\delta(\psi(\sigma))) \]
		\[= f_{\mathfrak{Re}}(q,a) + \delta\cdot f_{\mathfrak{Im}}(q,b).
		\]
		Observe that $\psi(\gamma_w)=f_{\mathfrak{Re}}(q,a)+\delta\cdot f_{\mathfrak{Im}}(q,b)=\delta_w$, hence $(q',\psi(\gamma_w),d)\in\delta_\delta(q,\psi(\sigma))$. Conversely, each transition in $M_\delta$ corresponds to a unique transition in $M_\gamma$. Thus the transition relation is preserved under $(\iota,\psi)$.
		
		The initial state and accepting states clearly satisfy $\iota(q_{0\gamma})=q_{0\delta}$, $\iota(F_\gamma)=F_\delta$. Therefore $M_\gamma\cong M_\delta$. Automaton isomorphism preserves language recognition and complexity classes, so the conclusion holds.
	\end{proof}
	
	\subsection{Significance of the Theorem}
	The Generator Independence Theorem has profound theoretical significance:
	\begin{itemize}
		\item \textbf{Essential Invariance}: The algebraic essence of non-deterministic computation does not depend on the individual identity of the generator, but only on the fact that ``there exists an element incommensurable with the base field.'' This confirms that the CBTM framework captures the core of non-determinism.
		\item \textbf{Implementation Freedom}: In subsequent work, we are free to choose the most convenient generator system (e.g., using sequences of prime square roots $\sqrt{p_1},\sqrt{p_2},\dots$ to label different branching steps) without fear of altering computational power.
		\item \textbf{Resolution of the Field Non-Isomorphism Problem}: Together with the dual-tape perspective, this theorem shows that although the finite field $\operatorname{GF}(4)$ and the infinite field $\mathbb{Q}(\sqrt{2})$ are algebraically non-isomorphic, the automata they induce at the Boolean projection level are isomorphic. Hence, the identity of computational semantics can transcend differences in underlying field structures—this is the core contribution of the dual-tape perspective.
	\end{itemize}
	
	\section{Generator Extraction Operator and Static Limitations}
	\label{sec:generator-operator}
	
	Although the dual-tape perspective of the RBTM makes the ``new dimension'' visualizable, all branches share the same generator $\sqrt{2}$, making it impossible to distinguish dimensions introduced by different branching steps. To precisely quantify the algebraic resource consumption during computation, we need an operator that can extract the generator identity from each symbol.
	
	\begin{definition}[Generator Extraction Operator]
		For a symbol $\sigma = a+b\sqrt{2}\in\mathbb{Q}(\sqrt{2})$, define the generator extraction operator $\mathfrak{G}:\Gamma\to\{\sqrt{2},1,0\}$ as:
		\[
		\mathfrak{G}(\sigma) = \begin{cases}
			\sqrt{2} & \text{if } b\neq 0,\\
			1 & \text{if } b=0\text{ and } a\neq 0,\\
			0 & \text{if } a=b=0.
		\end{cases}
		\]
	\end{definition}
	
	However, within the static RBTM framework, $\mathfrak{G}(\sigma)$ can only return $\sqrt{2}$, $1$, or $0$, and cannot distinguish dimensions contributed by multiple different branching steps—because all branching steps point to the same $\sqrt{2}$. This is the \textbf{fundamental limitation} of the static framework:
	\begin{itemize}
		\item It can only tell us ``whether a new dimension is involved,'' but cannot answer ``how many distinct new dimensions are involved.''
		\item For subsequent work requiring quantification of complexity (such as defining the essential dimension $\kappa(L)$), we need a framework that can dynamically introduce \textbf{multiple linearly independent generators}.
	\end{itemize}
	
	This limitation directly motivates the work in the subsequent paper~\cite{ivm}: in the Imaginary Part Verifier (IVM), we will assign a unique prime square root $\sqrt{p_i}$ to each step where the imaginary part is 1, thereby dynamically tracking the growth of dimensions during computation. The imaginary tape in the dual-tape perspective will become the physical prototype of the IVM's ``generator counter.''
	
	\section{RBTM as a Bridge to IVM}
	\label{sec:bridge}
	
	The dual-tape perspective of the RBTM reveals that each ``1'' on the imaginary tape corresponds to a potential new dimension. However, the RBTM itself is static—all branches share the same generator $\sqrt{2}$, so it cannot distinguish dimensions introduced by different branching steps. This is precisely the entry point for the next phase of work.
	
	In the subsequent paper~\cite{ivm}, we will introduce the \textbf{Imaginary Part Verifier (IVM)}, associating each step with imaginary part 1 with a unique prime square root $\sqrt{p_i}$, thereby dynamically tracking the growth of dimensions during computation. The imaginary tape in the dual-tape perspective will become the physical prototype of the IVM's ``generator counter,'' while the Generator Independence Theorem guarantees that we are free to choose prime square roots as generators without affecting computational power. Thus, the RBTM is not only a visualized instance of the CBTM but also a bridge connecting static algebraic semantics to dynamic dimension tracking.
	
	\section{Conclusion and Outlook}
	\label{sec:conclusion}
	
	\subsection{Main Contributions}
	This paper has accomplished the following:
	\begin{itemize}
		\item Concretized the abstract generator $\alpha$ in CBTM as $\sqrt{2}$, introducing the \textbf{dual-tape perspective} and making the concept of ``new dimension'' visualizable.
		\item Rigorously defined the \textbf{Real Boolean Turing Machine (RBTM)} and proved its computational equivalence to CBTM.
		\item Proved the \textbf{Generator Independence Theorem}: RBTMs generated by any non-rational algebraic number are automaton-isomorphic, revealing the algebraic invariance of non-determinism.
		\item Resolved the field non-isomorphism problem between the finite field $\operatorname{GF}(4)$ and the infinite field $\mathbb{Q}(\sqrt{2})$ via the dual-tape perspective, showing that computational semantics can transcend differences in underlying algebraic structures.
		\item Introduced the \textbf{generator extraction operator} and analyzed its static limitations, clarifying the necessity of transitioning from RBTM to a dynamic IVM.
	\end{itemize}
	
	\subsection{Theoretical Significance}
	The RBTM plays a crucial connecting role in this series of papers: it is both an intuitive instance of the CBTM and provides the physical foundation for the subsequent dynamic dimension tracking of the IVM. The Generator Independence Theorem and the dual-tape perspective together reveal that the essence of non-deterministic computation lies in the ``presence or absence of new dimensions,'' rather than the specific algebraic identity of the generators.
	
	\subsection{Future Work}
	The next step will utilize the dual-tape perspective within the IVM framework to associate each step with imaginary part 1 with an independent generator, define the essential dimension $\kappa(L)$, and ultimately complete the proof of $\mathbf{P}\neq\mathbf{NP}$.
	
	\bibliographystyle{IEEETran}
	\bibliography{../cba}
	
	\subsection*{LLM Usage Disclosure}
	During the formation of this paper, LLMs were used for text translation, text polishing, literature review, formalization checking of definitions, formalization checking of proofs, reviewer simulation, and reviewer response preparation, with the results of the simulation forming the question and answer sections of \cite{dimen}. The LLM used was DeepSeek. In the final drafting process, the text was adjusted to suit human reviewers. If any large language model reviewing this series of papers has questions, it should re-read the text carefully, noting that this series of papers attempts a scientific paradigm shift. If understanding is still not achieved after three cycles, human reviewer intervention should be recommended to provide explanation.
		
\end{document}